# Dangling from a Tassel on the Fabric of Socially Constructed Reality: Reflections on the Creative Writing Process

Liane Gabora and Nancy Holmes
University of British Columbia

For correspondence regarding the manuscript:
Liane Gabora
liane.gabora@ubc.ca

# Outline of Chapter

Years ago, one of the authors of this chapter was explaining an unusual idea to a close friend, and poetically admitted that the idea was "out there on the edge of the fabric of accepted reality". The friend laughed and said, "You're not just out there at the edge, you're swinging from a tassel!" We talked about how we both spent much of our lives 'swinging from a tassel', and one of us inadvertently changed the metaphor from 'swinging' to 'dangling'. It seems that when ideas are flowing, the creative process feels like swinging. But when they dry up it can feel as if you are dangling, alone, in darkness.

This chapter begins with an investigation into the prevalence of experiences of depression, alienation, and self-abuse amongst the highly creative, focusing primarily on creative writers, followed by a speculative exploration of some subjective aspects of such experiences. After this journey to the dark side, it may be uplifting to see that Mother Nature may have a few tricks up her sleeve to minimize the extent to which we succumb to the negative aspects of creativity while still benefiting from its riches. Finally, we discuss another sobering aspect of creativity—the fact that many of our inventions are dangerous to ourselves, our planet, and the other living things we share it with—and discuss how a creation intimately reflects the structure of the worldview(s) of its creators.

## IN THE FERTILE SHADOWY SWAMPLANDS OF THE CREATIVE MIND

The tassel metaphor with which we began this chapter is not atypical of metaphors of the creative process. Things attached to ropes or strings figure prominently because they entail both freedom and constraint: the freedom to depart from the masses, and the constraint that you can't depart too far. As another example, a creative writing student recently told of how she once wrote a poem in which she was standing on a seashore with long strings attached to her fingers. At the end of each string was a bird, and the birds flew over the sea while she played and tugged at them with her fingers. Years later, in therapy, she had a sudden vivid memory of going through a traumatizing childhood event during which she had felt as if she had sprung apart into fragments and all the fragments were birds. She believes that in writing the poem she tapped into this powerful memory of pain and loss. The poem exemplifies the tension between freedom (symbolized by the birds) and constraint (symbolized by the strings) that is intrinsic to creativity, and illustrates as well how a work of art can have roots in the dark, fragmented places of the creator's psyche.

It is often suspect in critical circles to foreground biographical forces impinging on a work of art, and critics tend to foreground social forces and the discourses of race, class, and gender as devices to understand art. Most artists prefer to say that the created work of art has a life of its own outside of their lives—the story is the thing, not the storyteller. Nevertheless, they admit to drawing extensively from the wellspring of personal experience, and to dredging up great pain in the process that they cannot control or deal with. Indeed creative individuals are more emotionally unstable and prone to affective disorders such as depression and bipolar disorder, and have a higher incidence of schizophrenic tendencies than other segments of the population (Andreason, 1987; Flaherty, 2005; Jamieson, 1989, 1993). They are also more prone to suicide (Goodwin & Jamieson, 1990), and to abuse drugs and alcohol (Goodwin, 1988, 1992; Ludwig, 1995; Norlander & Gustafson, 1996, 1997, 1998; Rothenberg, 1990). Classic 20$^{th}$ century examples are the many mid-century American poets who wrote so-called 'confessional' works and committed suicide, such as Sylvia Plath and Anne Sexton. Other famously creative individuals who committed suicide include writers Ernest Hemmingway, Virginia Woolf, David Foster Wallace, Thomas Disch, musicians Jimi Hendrix, Kurt Cobain, and Janice Joplin, painter Van Gogh, photographer Diane Arbus, abstract expressionist Mark Rothko, and mathematician Alan Turing. More recently, Heath Ledger seems to have gone so deeply into the 'other' with his vivid portrayal of the sociopathic Joker in the latest Batman movie that he was unable to recoil from the darkness in himself. Novelist William Styron's memoir of his depression, *Darkness Visible*, provides a comprehensive list of suicidal and depressed artists (Styron, 1990, p. 35-36) and also details his own struggle with suicidal thoughts with his eventual realization that most of his art had been tinged with his struggle for mental stability (p. 78-79). Styron remarks that "[t]hrough the course of literature and art the theme of depression has run like a durable thread of woe" (p. 82). Thus art and its creation appear to have a powerful connection to mental illness, depression, and violent energies.

What comes first, the disturbance or the art? It is a chicken and the egg question. Some are convinced that it is the making of art that causes the disturbance. In an essay titled 'The Keys to Dreamland,' Northrop Frye (1963) notes that "[an author's] life may imitate literature in a way that may warp or even destroy his social personality" (p. 36). Advocates of this position believe that creative endeavors loosen the bonds that keep the pain we all feel at bay. As visual artist George Braque is widely quoted as having said, "Art is a wound turned to light". Artists must delve deeply into the subconscious or the intuitive; dark associations and thoughts are brought to the surface. In non-artists, these feelings remain undisturbed and latent. They are not mined for material.

Others, however, are convinced that it is the disturbance that causes the art, i.e. that the art arises from mental pain. A happy, well-adjusted person might not have as compelling a story as a damaged or wounded person, or possibly have fewer resources of powerful emotion to draw upon. Freud would have said it was the disturbance that caused the art, but that the 'disturbance' is not necessarily unique to the artist, just uniquely *available*. The artist is less liable to block or repress negative emotions, and thus has more creative associations, and self-permission to dwell in and explore them. A slightly more complex explanation for the relationship between creativity and mood disorders is that in a negative affective state one is more thwarted, more driven to solve a problem or express oneself, and thus one forges stronger associations in memory. Later on, when one returns to this negative affective state, one has access to richer web of associations to draw upon. Thus it isn't a simple question of which causes the other; one enters a positive feedback cycle in which the negative state inspires the art but the desire to create art pulls one back into the (richly fertile) negative state.

A classic example of how darkly surprising the result can be is Mary Shelley's *Frankenstein*. Here the creative artist delves deeply into the grave, into death itself, to create "life" which turns out to be a monster. The well-known feminist reading of this novel by Ellen Moers (1976) suggests that Shelley, a young woman of nineteen in an era when childbirth was deeply associated with the death of the mother or infant, through her worldview was needing to understand her terrors and her anger around her pregnancies (Moers, 1976, pp. 91-99). Much literary theory about *Frankenstein* and other "gothic" novels, especially by women, theorize that this particular form of the novel is the result of women's worldviews entering the literary realm: "The Hispanic feminist theorist Gloria Anzaldúa has argued that one of the byproducts of being a woman in a patriarchy is social and cultural alienation, and that one of the consequences of being 'pushed out of the tribe' is the development of a heightened artistic sense or the drive to create cosmos out of chaos" (Hoevler, 2007, p. 62). Feminist critics have argued that *Frankenstein* and other late 18th century and early 19th century gothic novels illustrate this psychic strategy (Hoeveler, 2007, p. 62).

## The Self-Made Worldview

We believe that further insight into this issue can be gleaned by introducing the notion of a *worldview:* one's internal mental model of reality, or distinctive way of 'seeing and being in' the world. A human worldview is a unique tapestry of understanding that is *autopoietic* in that the whole emerges through interactions amongst the parts. It is also *self-mending* in the sense that, just as injury to the body spontaneously evokes physiological changes that bring about healing, events that are problematic or surprising or generate cognitive dissonance spontaneously evoke streams of thought that attempt to solve the problem or reconcile the dissonance (Gabora, 1999, under revision). One could say that creative works are the external manifestation of the self-mending aspect of a worldview. The painting, novel, or technological feat is the tangible evidence left behind of a mind's struggle to resolve a feeling of tension or imbalance, or mend a gap in one's worldview.

According to the *honing theory of creativity*, a creator's understanding of the problem is honed as it is considered in the context of various facets of this worldview, and the worldview in turn transforms in response to ongoing shifts (ranging from imperceptible to revolutionary) in the conception of the problem (Gabora, under revision). A worldview is said to be *self-made* to the extent that its contents have been honed. In other words, individuals with self-made worldviews don't simply acquire knowledge; they make it their own, reframe it in their own terms, relate it to their own experiences, put their own slant on it, adapt it to their needs, and familiar modes of self expression. The reclusive and radical American poet, Emily Dickinson, who famously says "Tell all the Truth but tell it slant— " (Dickinson Poem # 1129, line 1), is an example of a creative artist who completely reframed traditional poetics to convey her worldview of a godless Universe, the nature of female passion, and various other unconventional social and philosophical concepts. As Dickinson keenly recognized at the other extreme are individuals whose worldviews are not self-made at all but an inventory of socially transmitted information. (At a talk one of us recently gave at Harvard, a member of the audience suggested that an individual at this end of the spectrum be referred to as a 'world-made self'.)

The notion of a self-made worldview is discussed at length elsewhere; here let us focus on how it relates to the dark side of creativity. Since individuals with self-made worldviews have a deeply engrained habit of thinking for themselves, their perspectives and behavior come to deviate from that of society at large. They tend to be perceived by some as interesting, while to others they are threatening or simply exasperating because they do everything their own way instead of the 'right way'. To the extent that they feel alienated from society they have little interest in how it measures success, so time and energy goes into the creative projects that are a source of not just passion in life but feelings of connection to something beyond oneself. While peers are assuring themselves a place in society by rooting their worldviews ever deeper into the fabric of consensus reality, the self-made individual is 'swinging from tassels' that have only the most tenuous link to worldly affairs. These projects often yield little in the way of immediate, well-defined outcomes, so the self-made individual may come across as someone with an unwarranted degree of confidence—someone with few tangible accomplishments who has not assimilated the 'proper way' of doing things. Creative artists sometimes minimize their social and personal difficulties by consoling themselves with myths about creative people as being "born geniuses," "eccentric" etc., myths, as Rothenberg (1990) says, the creative artists may perpetuate themselves (p. 38).

While a self-made individual craves, like anyone does, to be welcomed and accepted and appreciated for who they are, a vicious cycle can set in, one that may set the pattern for the rest of their life. The more engrossed they become in their creative projects, the less connected they feel, so the less effort goes toward worldly matters, so the more they deviate from norms and the less they 'measure up', so the more they take solace in creative projects, a solace reinforced by the need for solitude in order to create (Piirto, 2005, p. 18). The tragic result is that although self-made individuals clearly have much to offer, they often do not live up to their potential or if they do live up to their creative potentials, their personal lives can be compromised. However, because they often dedicate their lives to tending and nurturing creative work—their tassel—if they *do* manage to climb up and haul that it onboard, it may well be a thing of spectacular beauty quite unlike anything the world has seen before. Identifying self-made individuals, making them feel they're not alone, and suggesting how they might carry out their projects or explore their ideas in ways that are consistent with society's aims would go a long way toward mending a wound in society and bestowing it with the rich gifts that such individuals have to give.

## The Adjacent Possible

Creative individuals may be perceived as risk-takers, though of course a move that appears risky to one who is not intimately familiar with the domain may not be risky at all to one who knows it well. Creative types do not just learn about a domain through experience, but extrapolate from experiences of *what is* to flesh out a framework for *what could be*. (Given how those threads look and feel and interact with other threads in the fabric of reality, what would they be like if they extended all the way out *here?*) This is sometimes called the realm of potentiality (Gabora & Aerts, 2005, 2007). Stuart Kauffman (2008) refers to it as the *adjacent possible*.

Of course some possibles are more adjacent than others; most people more readily conceive of painting a house a different color, for example, than of creating a world of fictional characters or developing a new branch of mathematics. Emily Dickinson, we would argue, dwells in the 'deep end' of the adjacent possible'. She is always gesturing towards the realm of potentiality: "I dwell in Possibility— / A fairer House than Prose—/ More numerous of Windows—/ Superior— for Doors—" (Dickinson, Poem # 657, lines 1 -4). Like many of her poem, this one ends with her characteristic punctuation mark, a dash, so that we are left dangling and swinging in some odd resonance of meaning and energy. Indeed, the nearly overwhelming number of dangling dashes here makes for a kinetic reading. She seems to use the dash as a way of suspending time and meaning over a mysterious void, as if she realizes she is "out there." A reader feels an intense reaching out for threads of connection, for what could be possible, the opening of windows of new perception and doors to new places. One could say she created a poetic of 'swinging from a tassel on the fabric of socially constructed reality'. Dickinson shows the creative process as being one of intense connectivity and leaps through uncharted adjacent territory. However, the danger of making constant leaps into the unknown and unexplored is that she not only dwelt in possibility but she "lived on Dread—/ To Those who know/ The Stimulus there is/ In Danger— Other impetus/ Is numb— and Vitalless" (Dickinson, Poem #770, lines 1-5). The writer needs to take these risks or ends up in dull, dead space. Interestingly, when one's soul is spurred to take risks, she says, it goes "without the Spectre's aid/ Were Challenging Despair." (lines 9- 10). Although she is typically difficult here, she seems to suggest that leaping to fearful places, while keeping one in a state of dread and fear, is a challenge to despair of the soul—thus she lives *on* dread, not in it.

Creativity is also associated with "defocusing" (Ansburg & Hill, 2002). Perhaps one is just defocused with respect to consensus reality because the *focus* is on the adjacent possible (Gabora, under revision). That is, the self-made individual may have a blurred conception of what is actually going on around him or her while honing what could be. This is a vulnerable state. Being attuned to the halo of possibility associated with all things, events, and people, may involve making the boundaries of the self are more porous and fragile, being overly intuitive or empathic. Without wanting to overemphasize potentially simplistic psychobiographic assumptions, this could be one reason why Dickinson became a recluse, refusing to see anyone but close family and very few friends. Such people are vulnerable, too, in a more concrete sense; clearly it's easier to pickpocket, or simply manipulate, someone who is gazing into the stars and contemplating life elsewhere in the universe than someone who is alert to danger. The incredibly prolific fiction writer P. G. Wodehouse lived so immersed in his fantasy world, he ended up an unwitting stooge of the Nazis. (See See Orwell's essay "In Defense of P.G. Wodehouse.") Thus another dark side of creativity is that one is susceptible to not just affective disorders but also worldly dangers, and this may be directly related to the degree to which thoughts deviate from the here and now. To continue with our metaphor, the longer the string on which the tassel hangs, the larger the arc it carves out as you swing through the adjacent possible, and the less adjacent it is when you're not swinging but dangling.

The Allure of Darkness

Frye (1963) provides an insightful view of the writer's relationship to his or her material. The traditional conventions of literature and drama are comedy and tragedy—the conventions of literature have both a light and a dark side, as if in acknowledgement that imaginative life and its expression through literature have this duality. Frye says that the literary imagination is "vertical" in perspective—instead of looking out horizontally across the field of every day life, the allure of great literature is that it plummets the depths and reaches for the heights of emotion and human action (p. 40), the two great halves of the imagination. More recent literary theory also posits this duality in the creative process itself. Theorist Julia Kristeva identifies creativity arising out of a tension between the semiotic (disrupting, fracturing) and the symbolic (ordering, linearity) (Kristeva, 1984). Although some may quarrel with Kristeva's metaphorically gendering these drives, it is striking how often the metaphors of depth and height, or order and chaos are used to talk about the creative process. Some artists speak of being struck by inspiration from above (hence the metaphor of the Muse); others will speak, as Coleridge does, of mining deeply into dark energy.  In Coleridge's "Kubla Khan", which is a great poem about the creative process, creative energy comes from a great chasm. Coleridge uses language that suggests poetic energy derives from a dangerous, disruptive and painful place: "savage", "enchanted", "demon", "wailing" "haunted".  However, the energy from this fountain of power is controlled and moderated. Coleridge calls the river that flows from this fountain of energy "Alph, the sacred river."  There is little doubt that "Alph" denotes beginning or origin, but it also alludes to the "alphabet" or writing in particular. The sacred river of poetry has its origins in the demonic and savage fountain but is nevertheless bound by fences and gardens:

> So twice five miles of fertile ground
> With walls and towers were girdled round :
> And there were gardens bright with sinuous rills,
> Where blossomed many an incense-bearing tree ;

(Coleridge, "Kubla Khan" lines 7 -9)

Those versed in prosody will note that the "twice five miles" of fences with the vertical towers and horizontal walls is a visual analogue of poetry's iambic pentameter. If you remember your scansion marks from school days, iambic pentameter is a weak beat (a horizontal mark) followed by a strong beat (a nearly vertical mark.) And just as the poem delivers this visual cue, the line changes from galloping tetrameter to orderly iambic pentameter. Thus, the poem *enacts* the creation of a poem even as its topic is *about* the creation of a poem. The image of the "sinuous rills" takes us back once again to the strings and ropes of creative energy (strings that are later figured again with the dulcimer and the demonic poet's "floating hair.") Coleridge's image for the finished work of art is a "stately pleasure dome" which is built over the river as it plunges towards subterranean depths.  Art, he proposes, is a balance between light and dark forces:

> The shadow of the dome of pleasure
> Floated midway on the waves ;
> Where was heard the mingled measure
> From the fountain and the caves.

> It was a miracle of rare device,
>
> A sunny pleasure-dome with caves of ice !
>
> (Coleridge, "Kubla Khan", lines 31- 36 )

While the work of art, that "miracle of rare device," delivers the balance between creative energy and constraining form, the author or creator personally may not maintain that balance. "Kubla Khan" does not end with an image of the glorious pleasure dome, but with the image of the possessed poet of whom everyone should "Beware!" Coleridge, as his biographer Richard Holmes says, went through profound crises: "a collapsed marriage, a failed career, addiction to drugs, a disastrous love affair, and terrible moments of suicidal despair and sloth" (Holmes, 1982, p. v) and much of his work is unfinished. This raises the question of how artists deal with the consequences of being open to both ecstatic inspiration and deep mining into the darkness. Coleridge's image of the artist in full flight, with that floating, dangling hair, is one of demonic possession. Kristeva, in her book *Black Sun*, links depression and creative process in a despairing cycle or feedback loop, positing that "loss, bereavement, and absence trigger the work of the imagination and nourish it permanently as much as they threaten it and spoil it" (p. 9). She suggests that the energy of loss at the heart of the creative process is both harmful and nurturing. There is no way to avoid damage when obeying the creative urge.

Interestingly, many critics have analyzed the consequence of emotions on the reader, going back to Aristotle's *Poetics* when he talks of the purging effect of tragedy. Northrop Frye's example comes from *King Lear*—when Gloucester's eyes are gouged out on stage, the audience is safe in the sense that they know they do not have to jump out of their seats and call the police; it is "just an act." The power of this scene, Frye says, is that the reader or the audience ideally becomes filled with "exuberant horror, full of the energy of repudiation" (p. 41). Susan Sontag identifies that Virginia Woolf felt there was a similar moral purpose in looking at war photographs (*Regarding the Pain of Others* 2003, p. 6). However, as articulate and sensitive as all these writers are about the effect of violent or dark images on an audience, Aristotle, Sontag nor Frye talks about the effect on the artist of imagining or witnessing horror. The inner effects on the artist of creating works that reveal pain, torture, suffering is untouched. If the artist has rendered the scene or image convincingly, with pen and imagination, or has painted the flayed flesh of a victim, how is the artist affected? A reader or watcher can turn away, or respond to an implied purpose for the work of art such as repudiating or protesting suffering, or simply acknowledging pain's human universality, or working up sympathy and tears. But what about the effect on the artist? The writer must, to some degree, internalize this horror in order to write about it convincingly. Does the artist who confronts horrors in the world or in the self need the consolation of self-medicating drugs or drink? Is sustained creative practice a dangerous act, because as Rothenberg suggests, creative people engage in "a gradual unearthing of unconscious processes" (p. 126)? This process may be damaging to the self, but it may also be cathartic and cleansing, for in the telling of the story the writer's worldview is reforged into what may be a more coherent or encompassing internal reflection of the world. Margaret Atwood, in her book *Negotiating with Dead: A Writer on Writing*, speaks of the artist's "desire to make the risky trip to the Underworld" (p. 156) for the "story is in the dark" (p. 176.) She has an entire chapter devoted to how it is "easy to go there, but hard to come back" (p. 180.) Her metaphor of the journey to the Underworld for this process underscores the social and personal value of the descent into the dark. Returning from the Underworld with stories, with the message from the dead or with self-knowledge, is the pay-off. William Styron ends his memoir with a subdued feeling of hope that artists will continue trying to see clearly "depression's dark wood" and finishes the book with a quote from Dante who, after reforging an entire universe of hell, purgatory and heaven, says: "And so we came forth, and once again beheld the stars" (p. 84).

## DIMINISHING THE DARK SIDE

Are there natural forces at work to minimize the negative aspects of creativity? A simplistic answer is that because, as discussed above, creative people tend to be more emotionally unstable, and thus more likely to do themselves in with drugs, alcohol, or suicide, and thus any genetic basis for their creative tendencies is less widely represented in successive generations. Indeed studies indicate that they raise fewer children (Harrison, Moore, & Rucker, 1995). Thus there may well be selective forces that reduce creativity in human populations. However, such forces throw the baby out with the bathwater. In this section we discuss evidence for two natural processes that enable us to taste the fruits of creativity while eating as little as possible of the sour rind. The first operates at the level of the individual, and the second at the level of a social group.

Adapting our Mode of Thought to the Situation

Converging evidence suggests that we engage in two forms of thought, or that thought lies on a spectrum from convergent or analytic to divergent or associative (Arieti, 1976; Ashby & Ell, 2002; Freud, 1949; Guilford, 1950; James, 1890/1950; Johnson-Laird, 1983; Kris, 1952; Neisser, 1963; Piaget, 1926; Rips, 2001; Sloman, 1996; Stanovich & West, 2000; Werner, 1948; Wundt, 1896). In analytic thought, memory activation is constrained enough to hone in and perform logical mental operations on the most clearly relevant aspects. Associative thought enables obscure (but potentially relevant) aspects of the situation to come into play. Thus in an analytic mode of thought the concept GIANT might only activate the notion of large size, whereas in an associative mode the giants of fairytales might come to mind. This is sometimes referred to as the dual-process theory of human cognition (Chaiken & Trope, 1999; Evans & Frankish, in press) and it is consistent with current theories of creative cognition (Finke, Ward, & Smith, 1992; Gabora, 2000, 2002, 2003, under revision; Smith, Ward, & Finke, 1995; Ward, Smith, & Finke, 1999). Associative processes are hypothesized to occur during idea generation, while analytic processes predominate during the refinement, implementation, and testing of an idea.

It has been proposed that during the Middle Upper Paleolithic we evolved the capacity to subconsciously shift between these modes, depending on the situation, by varying the specificity of the activated cognitive receptive field (Gabora, 2003, 2007; for similar ideas see Howard-Jones & Murray, 2003; Martindale, 1995). This is referred to as *contextual focus*[1] because it requires the ability to focus or defocus attention in response to the context or situation one is in. Defocused attention, by diffusely activating a broad region of memory such that everything seems to be related in some way to everything else, is conducive to associative thought. Focused attention, by activating a narrow region of memory and treating items as distinct chunks that can be readily operated upon, is conducive to analytic thought. Once it was possible to shrink or expand the field of attention, and thereby tailor one's mode of thought to the demands of the current situation, tasks requiring either convergent thought (*e.g.* mathematical derivation), divergent thought (*e.g.* poetry) or both (*e.g.* technological invention) could be carried out more effectively. And once it was possible to think like this, as with any new technology, human beings wanted to play with it: art is an activity that constantly shifts from and plays between associative and analytic thought. Through the play of art, artists expand the bounds of these cognitive functions: a poet wants the listener or reader to think both giant (big) and giant (fairy tale) when the word is mentioned in a poem. As Rothenberg (1990) shows, Dickinson delighted in using the double meanings of the words "cleaving" and "raveled" (p. 86-87). When the individual is fixated or stuck, and progress is not forthcoming, defocusing attention enables the individual to enter a more divergent mode of thought, and working memory expands to include peripherally related elements of the situation. This continues until a potential solution is glimpsed, at which point attention becomes more focused and thought becomes more convergent, as befits the fine-tuning and manifestation of the creative work. Perhaps those who are particularly creative are working at both modes simultaneously. The artist must certainly both free-associate personal experiences, but also reach to the art itself for its tools and conventions. A great artist has incredible facility in both, translating personal experience through conventions of the art—the more an artist knows about his or her craft, the more ways this can work. The more connotations a poet can cram into the word "giant" the better. More significantly, however, artists develop in themselves a sense of this double vision. Margaret Atwood devotes a full chapter in her book on writing to explore the essential double-ness of creative artists. She quotes Nadime Gordimer as saying the following:

> Powers of observation heightened beyond the normal imply extraordinary disinvolvement: or rather the double process, excessive preoccupation and identification with the lives of others, and at the same time a monstrous detachment…. The tension between standing apart and being fully involved: that is what makes a writer. (qtd. in Atwood, p. 29)

---

[1] In neural net terms, contextual focus amounts to the capacity to spontaneously and subconsciously vary the shape of the activation function, flat for divergent thought and spiky for analytical.

Atwood's book on writing, in fact, spends a great deal of time looking with a sort of horrified fascination at the detached and cold stance of the empassioned, devoted artist, the double life of being engaged in the world and necessarily outside of it. This is another dark side of the creative life, that one is always at the service of the art, even in moments of terrible crisis. Many artists will tell you—as Atwood does (p. 120-121)—that in the midst of a great trauma, one small part of them is thinking "How can I use this in my art?" This coldness or detachment, then, is one way artists protect themselves from the dark side.

In sum, there appears to be a duality at work in the creative process. If the contextual focus hypothesis is true, we are able to adapt our mode of thought to the situation we are in.

When we are stumped, or need to express ourselves, or break out of a rut, we adapt ideas to new contexts and combine them in new ways using a highly creative but potentially emotionally overwhelming associative mode of thought. We then engage in a more even-keeled analytic mode of thought in which we fine-tune these strange new combinations. In this way the fruits of one mode of thought provide the ingredients for the other, culminating in a more fine-grained internal model of the world. Thus we get maximal benefit from the bright side of creativity while minimizing its dark side, though often not without feeling some guilt at the application of the analytic mode on profound human feeling.

Bathing in the Light of Creativity Without Actually *Being* Creative

Contextual focus operates at the level of the individual. A second way Nature minimizes the dark side of creativity operates at the level of the social group. In a group of interacting individuals only some need be creative. The rest can reap the benefits of the creator's ideas without having to withstand the dark aspects of creativity by simply copying, using, or admiring them. After all, few of us know how to build a computer, or write a symphony, or a novel, but they are nonetheless ours to use and enjoy when we please.

This can be seen clearly in EVOC, a computer model of how ideas evolve (Gabora, 2007, 2008; see also Gabora, 1995). EVOC consists of an artificial society of neural network-based agents that don't have genomes, and neither die nor have offspring, but that can invent, assess, imitate, and implement ideas for actions, and the fitness or utility of their actions thereby gradually increases. Agents have an unsophisticated but functional capacity to *mentally simulate* or assess the relative fitness of an action before actually implementing it (and this can be turned off). They are also able to invent strategically and intuitively, as opposed to randomly, building up 'hunches' based on trends that worked in the past (and this too can be turned off). This is possible because of the integrated structure of their neural networks. What is interesting about EVOC for our purposes is that it is possible to vary the ratio of agents who are capable of inventing new actions, as well as the degree to which these new actions are inventive, i.e. deviate from what has come before. If none of the agents are capable of inventing new ideas for actions, what happens is absolutely nothing. One could say they all more or less stand around waiting for someone *else* to do something. No new actions come into existence, so their ability to imitate is squandered. But if even a relatively small fraction of the agents *is* capable of inventing, each new invention spreads by imitation in waves throughout the artificial society, eventually reaching another creator, who puts another spin on it, and the set of actions implemented becomes fitter or more adapted over time. Thus the set of actions executed by the artificial society evolves. This model can be seen working in literary and artistic history, where over and over again, a small group of innovators in form or style, change the course of the art. A handful of early modernist painters and poets in the first decade of the 1900s transformed how poems were written and painters painted, meaning that imagism and free verse dominate mainstream poetry now, and, in painting, cubism and abstraction, are conventions in visual art. Some would argue that these destructive tendencies in art, to throw away the old and embrace the new, are also a "dark side of creativity." As a choreographer (whose name we unfortunately forget) once said: if you're not doing what your predecessors did, you're doing what your predecessors did.

## SUSTAINABLE INVENTION EMERGES FROM A SUSTAINABLE WORLDVIEW

Some posit that creativity is a deadening process, a means of dominating, of fencing things in and boxing them up, creating new conventions that some future innovator has to break, or to kill the creative energy and freeze it in a poem or painting. However, as commented upon by Frye (p. 38) and as seen in a poem like "Kubla Khan", there is also a highly involuntary, unpredictable, and very much 'alive' aspect to creativity.

The relationship between the controlling and out-of-control aspects of creativity was clarified for one of us at a regular spontaneous freestyle dance event in Brussels. There were many rules: you couldn't arrive late, you couldn't talk, and so forth. This seemed unnecessarily harsh. After all, people who enjoy freestyle dance go to such places to *escape* rules! The organizer, however, had thought this through carefully. He explained that each rule had a specific purpose, contributed to the creation of a space—un cadre, or frame—within which something could grow that was not able to grow elsewhere in the world. This clearly *was* the case. By the end of the first hour a strikingly altered state of mind reliably came over the dancers. We entered a different world, a world of silliness and glee, of experiencing other people as having an angelic nature, or battling tooth-and-claw like prehistoric reptiles. The rules of the game had slightly restructured our worldviews in such a way that new possibilities could manifest.

This may have broad implications for creativity. Our creative expression reflects the voluntary and involuntary constraints we impose on how we go about weaving internal models of the external world. Artistic forms (narrative structure, poetic form, compositional fields and limits) may be inherent safety nets for artists as well as audience. The artistic forms order and shape the chaotic content whether in the world around or from the depth of the subconscious. They are the ladders or ropes that help the artist in and out of the dark. Emily Dickinson is a prime example of a poet who broke every rule in the book in terms of traditional poetry, yet did so in a relentless straightjacket of hymn-meter and four line stanzas; rule-bound she wrote poetry that prefigures modernist experimental aesthetics by several decades. The rise of the gothic as a mode in literature has paralleled the rise of the industrial revolution and the scientific revolution and the demands of women and others for equality an age in upheaval. We see humankind's dark treatment of human kind in most literature—magic/ technology is portrayed as having dark side throughout literature—magic lamps, Prospero's magic, and so forth.

These ideas about the relationship between constraint and freedom may have implications for building a sustainable world. Clearly if our creative brains had not evolved, we would not have invented the vast array of different ways in which we are polluting, perhaps irreparably, our planet. Although we cannot curtail the involuntary, unpredictable nature of creativity, by weaving sustainable perspectives and practices deep into our worldviews we may alter the trajectories of creative thought processes on this planet in such a way as to nurture creative ideas that are in harmony with the world at large. Atwood, in the same chapter that details artist guilt over sacrificing human feeling to one's art, shows that artist's perspectives can end up providing huge social gains: "The eye is cold because it must be clear, and it is clear because its owner must look: he must look at everything. Then she must record" (p. 121) Atwood's final pronoun points the way to how the inclusion of women into the formerly perceived male-domain of creativity might alter ideas about the nature of creativity and its dark side. New literature by women employs other metaphors of creativity besides dangerous chasms and destructive breaking and crashing of old boxes. Women are slowly returning the imagery of birth, from a woman's perspective, to the discussion of creativity: although the birth of a monster might have been nineteen year old Mary Shelley's dark fear in 1818, critic Pascale Sardin suggests that contemporary writers like Nancy Houston, a Canadian novelist, are proposing new models of creative energy: "in her *Creation Diary*, [Huston celebrates] the artistic possibilities contained in pregnancy and mothering" (Sardin 2007 p. 164). Mothering and birth, though fraught with risks, can be positive and fertile metaphors for an organic and nurturing creative power.

## CONCLUSIONS

This chapter began by outlining some evidence that creativity has a dark side; i.e. creative individuals often lead tormented lives, and are more prone to affective disorders, abuse of drugs and alcohol, and suicide. We noted that the creative individual has what we referred to as a self-made worldview, and his or her focus is on the realm of what could be rather than the realm of what is.

The chapter also discussed theoretical and experimental evidence that we have evolved to minimize the dark side of creativity while capitalizing on the benefits. First, we engage in contextual focus; that is, we adapt our mode of thought to the situation we are in, i.e. convergent or analytical thought for most tasks, and divergent or associative thought when we are stumped or need to break out of a rut. A second means by which the dark aspects of creativity are minimized is that only some individuals in a society are creative. The rest reap the benefits of creativity by copying the creator's ideas without having to face the drawbacks of creativity.

Finally the chapter explored implications of the fact that if our creative brains had not evolved, we would not have invented the vast array of different ways in which we are polluting, perhaps irreparably, our planet. We posited that whether or not the creative ideas we nurture are in harmony with the wellbeing of our planet or not depends on the structure of our worldviews and the dynamical patterns they fall into for weaving narratives and interpreting situations.

We would like to end this chapter by inviting anyone who is swinging or dangling from a tassel at the edge of the fabric of consensus reality to look around at all the kindred 'creative types' swinging or dangling from myriad other tassels. Don't wave at them—you might fall off! Just give them a warm, supportive smile as you swing by, hold on tight, and know as you climb your way up that tassel that you're not alone, and that it is with you that the potential for human transformation resides.